\documentclass[12pt]{article}
\usepackage{graphicx}
\newcommand{\kmng}{K_{\mu 2\gamma}}
\newcommand{\kmnp}{K_{\mu 3}}

\title
 {
	{
		New search for T-violation in the decays of the charged kaon\\
	}
	~\\
	{
		\small\bf \underline{V.~Anisimovsky}\footnotemark[1], 
		A.~Ivashkin, Yu.~Kudenko\\
		~\\
		Institute for Nuclear Research of the 
		Russian Academy of Sciences,\\ 
		Moscow, 117312 ~Russia \\
		(For the KEK-PS E246 Collaboration)
		\footnotetext{$^{\ast}$E-mail: valera@al20.inr.troitsk.ru}
	}
}
\date{}
\begin{document}
\maketitle
\begin{abstract}
We report the results of the measurement of T-violating transverse muon 
polarization in the decays $K^+ \to \mu^+ \nu_\mu \pi^0$ ($\kmnp$) and 
$K^+ \to \mu^+ \nu_\mu \gamma$ ($\kmng$) performed in the experiment E246 
at KEK. Preliminary results obtained for the entire data set
taken in the period 1996-2000  are consistent with no T-violation 
in both decays.
\end{abstract}

\section{Introduction}

The purpose of the E246 experiment  is to measure the transverse 
component of the muon polarization in the decay $K^+ \to \mu^+ \nu_\mu \pi^0$ 
($\kmnp$), while we were also able to extract $K^+ \to \mu^+ \nu_\mu \gamma$ 
($\kmng$) decay as a by-product. The transverse muon 
polarization is a T-odd observable defined as $P_T=\vec{s}_\mu\cdot
(\vec{p}_{\pi(\gamma)}\times\vec{p}_\mu)/|\vec{p}_{\pi(\gamma)}\times\vec{p}_\mu|$
where $\vec{p}_\pi$ is used for $\kmnp$ and $\vec{p}_\gamma$ for $\kmng$, 
respectively. 

In the framework of the phenomenological consideration,  the transverse muon
polarization can be related to the $\kmnp$ and $\kmng$ form factors. 
For $\kmnp$ the T-violating polarization is proportional to the imaginary part 
of the ratio of $\kmnp$ form factors: $P_T \propto m_\mu m_K \mbox{Im}(\xi)$, 
where $\xi=f^-/f^+$ and $f^+$, $f^-$ are defined through 
\[
M_{\kmnp} \sim G_F \sin{\theta_c} \left[f^+(q^2)(p^\lambda_K+p^\lambda_\pi)+
f^-(q^2)(p^\lambda_K-p^\lambda_\pi)\right] 
\cdot \left[\bar{u}_\mu\gamma_\lambda(1-\gamma_5)u_\nu\right]
\] 
The Standard Model predicts a vanishing value of less than $10^{-7}$ for $P_T$ 
in $\kmnp$~\cite{pt_kmu3_sm}. The calculations of $P_T$ due to the electromagnetic 
final state interactions result in a 
value of  less than $10^{-5}$~\cite{pt_kmu3_fsi}. There are several 
non-standard models predicting sizeable value for $P_T$: multi-Higgs models, SUSY with 
squarks mixing, SUSY with R-parity violation, leptoquark models 
\cite{pt_kmu3_nsm,kobayashi}. The values predicted in these models vary from 
$4\times 10^{-4}$ to $10^{-2}$.

\begin{figure}[htb]
\includegraphics[width=7.0cm]{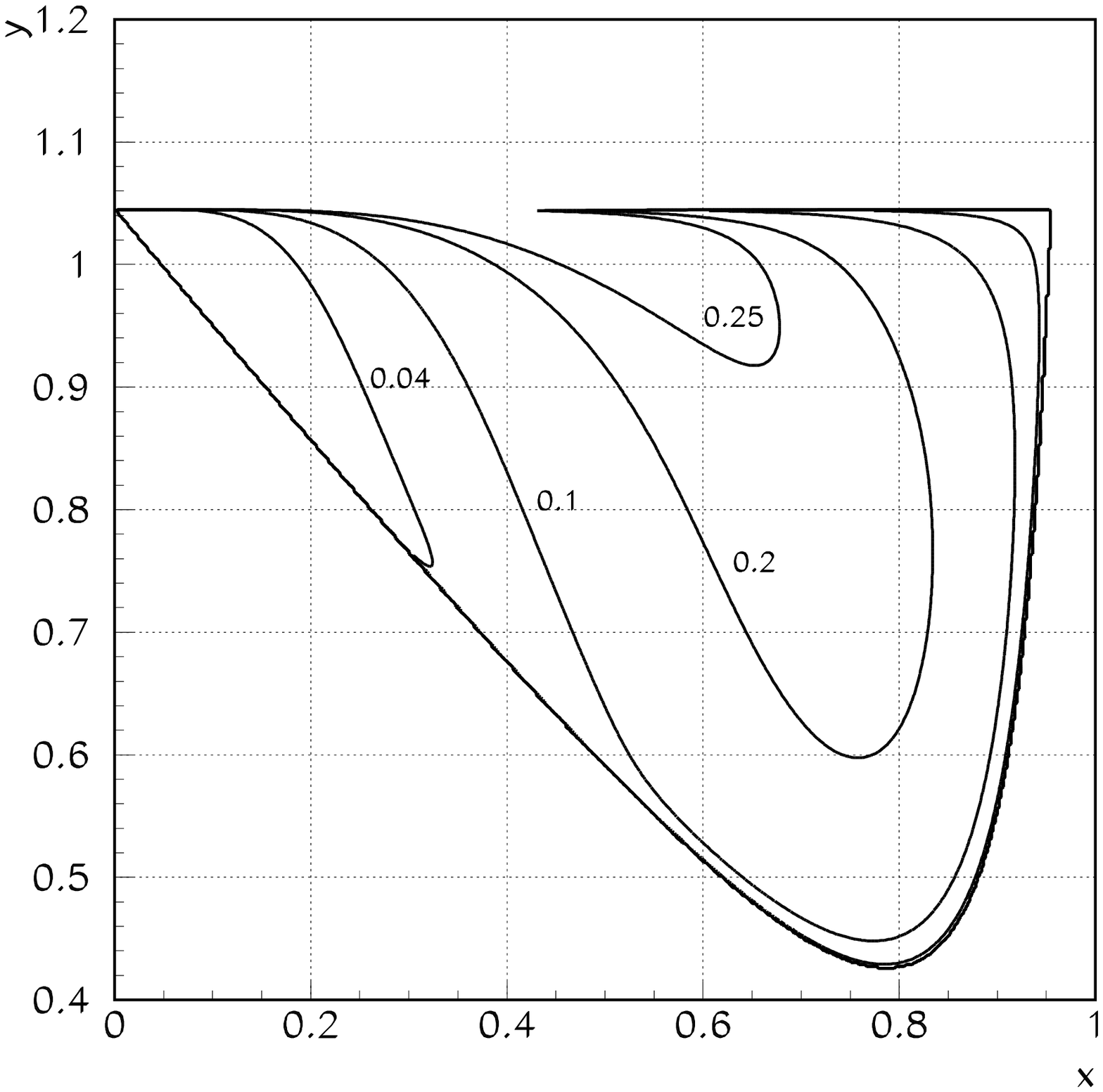} 
\includegraphics[width=7.0cm]{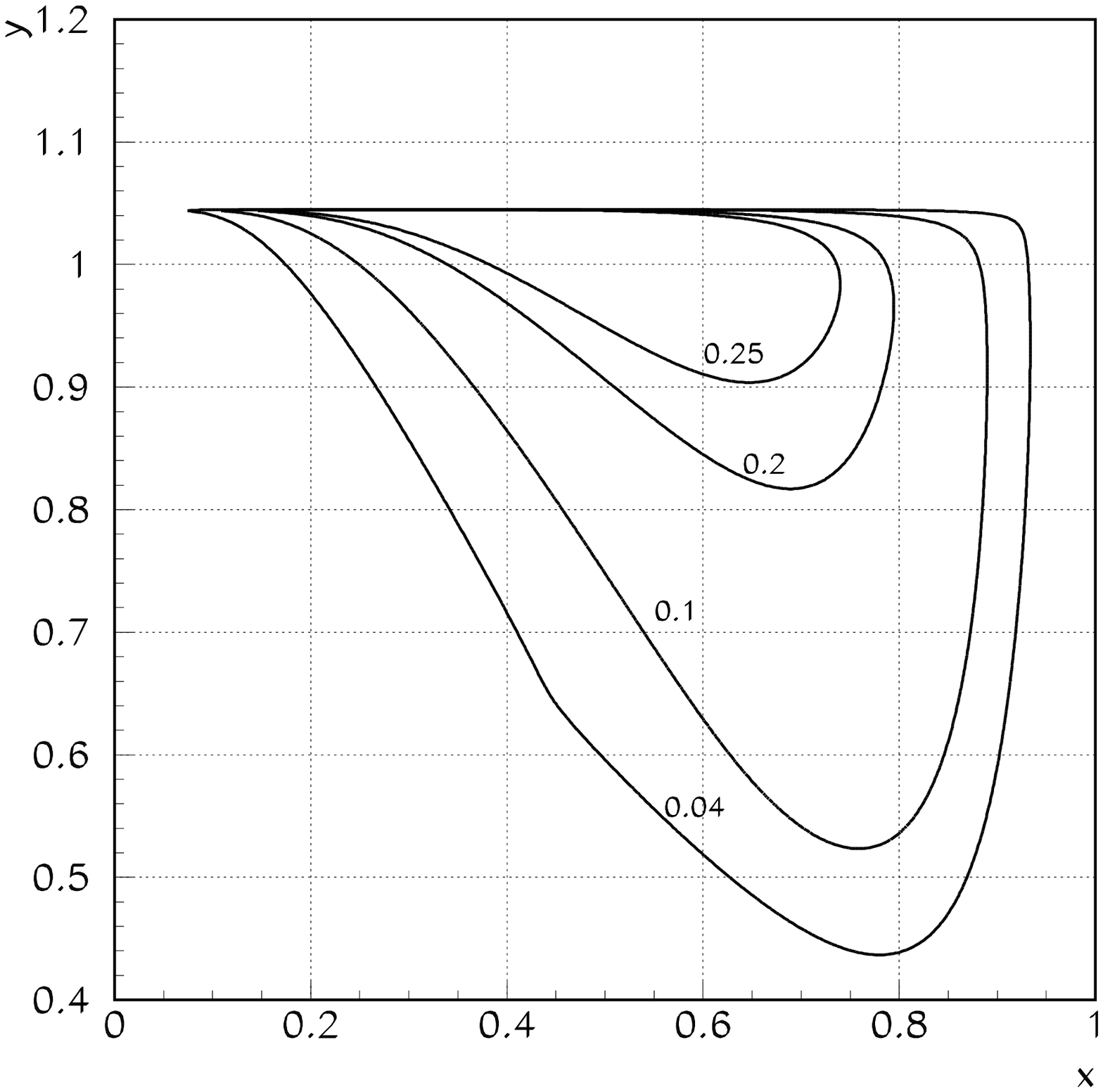} 
\caption{The contour lines  of $[\sigma_V(x,y)]$ (left) and 
$[\sigma_{V}(x,y)-\sigma_{A}(x,y)]$ (right) over the Dalitz plot. 
The standard $\kmng$ 
kinematic variables $x=\frac{2E_\gamma}{m_K}$ and $y=\frac{2E_\mu}{m_K}$.}
\label{fig:sigma}
\end{figure}

In case of $\kmng$ the transverse polarization is related to the decay
form factors in a more complicated way: 
$P_T(x,y)=\sigma_V(x,y)\mbox{Im}(\Delta_V+\Delta_A)+[\sigma_V(x,y)-\sigma_A(x,y)]\mbox{Im}(\Delta_P)$
where $\sigma_V(x,y)$ and $[\sigma_V(x,y)-\sigma_A(x,y)]$ are the functions 
of $\kmng$ kinematic parameters (shown in Fig.~\ref{fig:sigma}), 
and $\Delta_{(V,A,P)}$ are the contributions of non-standard interactions 
to the effective form factors~\cite{chen}.
Although the Standard Model prediction for $P_T$ in $\kmng$ is as small 
as for $\kmnp$~\cite{pt_kmu3_sm}, the contribution of the 
final state interaction to this value is $\leq 10^{-3}$~\cite{braguta}, i.e. 
 considerably larger than in the case of 
$K_{\mu3}$ decay.  The predictions for non-zero $P_T$ for $\kmng$ 
come from the same non-standard models mentioned for $\kmnp$ and also from 
left-right symmetric models~\cite{kobayashi,chen,wu2}. The expected values 
vary from $3\times 10^{-3}$ to $10^{-2}$.

The noteworthy peculiarity of these predictions obtained in 
different models is the correlations between 
the expected values of $P_T$ for $\kmnp$ and 
$\kmng$~\cite{pt_kmu3_nsm,kobayashi,wu2}:
in the three Higgs doublet model the $P_T$ expectations are related as 
$P_T(\kmnp)\sim 2 P_T(\kmng)$;
in SUSY with squarks mixing the relation  for $P_T$ induced by Higgs 
exchange: $P_T(\kmnp)\sim -2 P_T(\kmng)$, while for $P_T$ arising from 
W-boson exchange: 
$P_T(\kmnp) \sim 0$, $P_T(\kmng)\neq 0$;
in SUSY with R-parity violation the relation is $P_T(\kmnp)\sim P_T(\kmng)$;
in left-right symmetric models we have: $P_T(\kmnp)=0$, $P_T(\kmng)\neq 0$.

The previous experimental results for $\kmnp$ came from the BNL experiment 
which used kaon 
decays in flight~\cite{bnl}: $P_T=(-3.1 \pm 5.3) \times 10^{-3}$, 
$\mbox{Im}(\xi)=(-1.6 \pm 2.5) \times 10^{-2}$, as well as from E246~\cite{kek1}: 
the result obtained for the data collected during 1996-1997 period was 
$P_T=(-4.2 \pm 4.9(stat) \pm 0.9(syst)) \times 10^{-3}$, 
$\mbox{Im}(\xi)=(-1.3 \pm 1.6(stat) \pm 0.3(syst)) \times 10^{-2}$. Both 
 results indicated 
no T-violation in $\kmnp$. For $\kmng$ there has been no $P_T$ measurement, 
so our result 
is the first one.

\section{Experiment}

\begin{figure}[htb]
\centering\includegraphics[width=13.5cm]{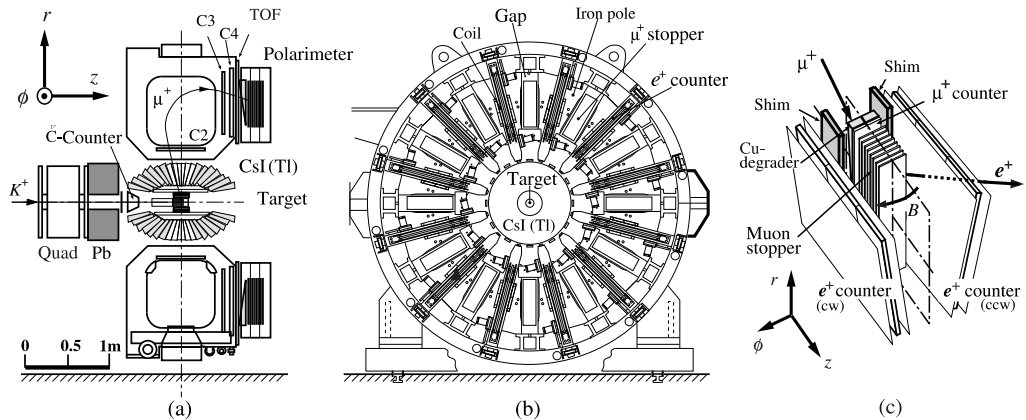} 
\caption{The layout of the  E246 detector: (a) side view, 
(b) end 
view and (c) one sector of the polarimeter.}
\label{fig:e246setup}
\end{figure}

The E246 apparatus  is shown in 
Fig.~\ref{fig:e246setup}, 
and described in detail elsewhere~\cite{e246setup}.
Kaons with $P_{K^+}=660$ MeV/c are identified by 
a ${\rm \check{C}}$erenkov counter, slowed in an Al+BeO degrader and then 
stopped in a target array of 
256 scintillating fibers located at the center of a 12-sector superconducting 
toroidal 
spectrometer. Charged particles from kaon decays in the target were tracked by 
means of multi-wire proportional chambers at the entrance (C2) and exit (C3 and C4) of 
each magnet sector, along with the target and a scintillation ring hodoscope 
around the target. The momentum resolution of $\sigma_p=2.6$ MeV/c at p=205 MeV/c 
was obtained using mono-energetic products from the two-body decay 
$K^+\to\pi^+\pi^0$ 
($K_{\pi 2}$). The energies and angles of the photons from $\pi^0$ decays 
were measured 
by a CsI(Tl) photon detector consisting of 768 modules. The photon detector 
covers a solid 
angle of 3$\pi$ steradian, with openings for the beam entry and exit and 
12 holes for 
charged particles to pass into the magnet gaps. To suppress accidental 
background from the beam, 
timing information from each crystal was used. A good time resolution of 3.5 nsec 
(rms) at 100 MeV was achieved. Energy resolution of 
$\sigma_E/E=2.7$\% at 200 MeV, 
angular resolutions of $\sigma_{\theta} = 2.3^{\circ}$ and the invariant mass 
resolution of $\sigma_{\gamma\gamma}=9$ MeV/c$^2$ were obtained.
Muons entering the polarimeter (Fig.~\ref{fig:e246setup}c) were 
degraded by an Al+Cu block and stopped in a stack of pure Al plates. 
Positrons from  $\mu^+\to e^+\nu\bar{\nu}$ decays of stopped muons were 
detected by positron counters which were located azimuthally between the 
 muon stoppers.

The trigger included the signals from ${\rm \check{C}}$erenkov, target, 
TOF and positron counters along with the requirement of at least 
one hit in CsI(Tl) calorimeter.

The T-violating asymmetry was extracted using a double ratio as:
\[
A_T =\frac{1}{4} \left[ \frac{(N_{cw} / N_{ccw})_{fwd}} {(N_{cw} / 
N_{ccw})_{bwd}} - 1 \right] \label{atdr}
\]
Here, $N_{cw}$ and $N_{ccw}$ are the sums over all 12 sectors  
of counts of 
clockwise ($cw$) and 
counter-clockwise 
($ccw$) emitted positrons and $fwd$/$bwd$ denote events with
the photon (or $\pi^0$) going forward/backward with respect to beam direction.
The sign of $A_{T}$ for $fwd$ events is opposite to that of $bwd$ events 
that allows us to  employ a 
double ratio method which reduces most systematic errors and enhances the 
effect. 
Moreover, considerable reduction of systematic effects was achieved due to 
the azimuthal symmetry of the 12-sector detector.

The value of $P_{T}$ is related to $A_T$ by
\[
P_T = \frac{A_T}{\alpha \times f \times (1-\beta)} \label{ptat}
\]
where $\alpha$ is the analyzing power of the polarimeter, 
$f$ is an angular 
attenuation factor and $\beta$ is the overall fraction of  
backgrounds.

\section{Analysis}

The extraction and analysis of $\kmnp$ and $\kmng$ events comprised several procedures 
common for both decays. The common stage included target analysis, charged 
particle 
tracking and TOF analysis and the analysis of muon decay in polarimeter.
Active target analysis included target energy deposition and target timing 
cuts to 
get rid of kaon decays in flight. The momentum of charged particle 
reconstructed 
by four-point tracking procedure was used to suppress $K_{\mu 2}$ and 
$K_{\pi 2}$ decays  
by selecting events with $p<190$ MeV/c. 
The cut on $\chi^2$ for the charged 
particle track was used to suppress $K_{\pi 2}$ decays 
with $\pi^+$ decay 
in flight. To separate muons and positrons 
(thereby suppressing $K_{e3}$) we 
used time-of-flight~(TOF) technique to calculate charged particle 
mass and then, on the 
scatter plot 
of the TOF energy deposition versus TOF-reconstructed mass square, we separated 
the 
muon cloud from the positron one. Finally, the common stage included the 
signal extraction
from the positron time spectra in the polarimeter.

The second stage of event selection was specific for each decay mode. For $\kmnp$ 
we selected one-photon events with $E_{\gamma}>70$ MeV and two-photon events 
which satisfy 
the constraints on the invariant mass of the two photons and on the missing 
mass (reconstructed mass of missing neutrino): 
$ 70<M_{\gamma\gamma}< 180$ 
MeV/c$^2$ and $-25000< M^2_{miss} < 20000$ MeV$^2$/c$^4$. Additionally, we 
used the cuts 
on the opening angles between two photons and between muon and $\pi^0$: 
$\Theta_{\gamma\gamma}>60^\circ$ and $\Theta_{\mu\pi}<160^\circ$ to suppress 
kaon decays in flight and $K_{\pi 2}$.

\begin{figure}[htb]
\includegraphics[width=7.0cm]{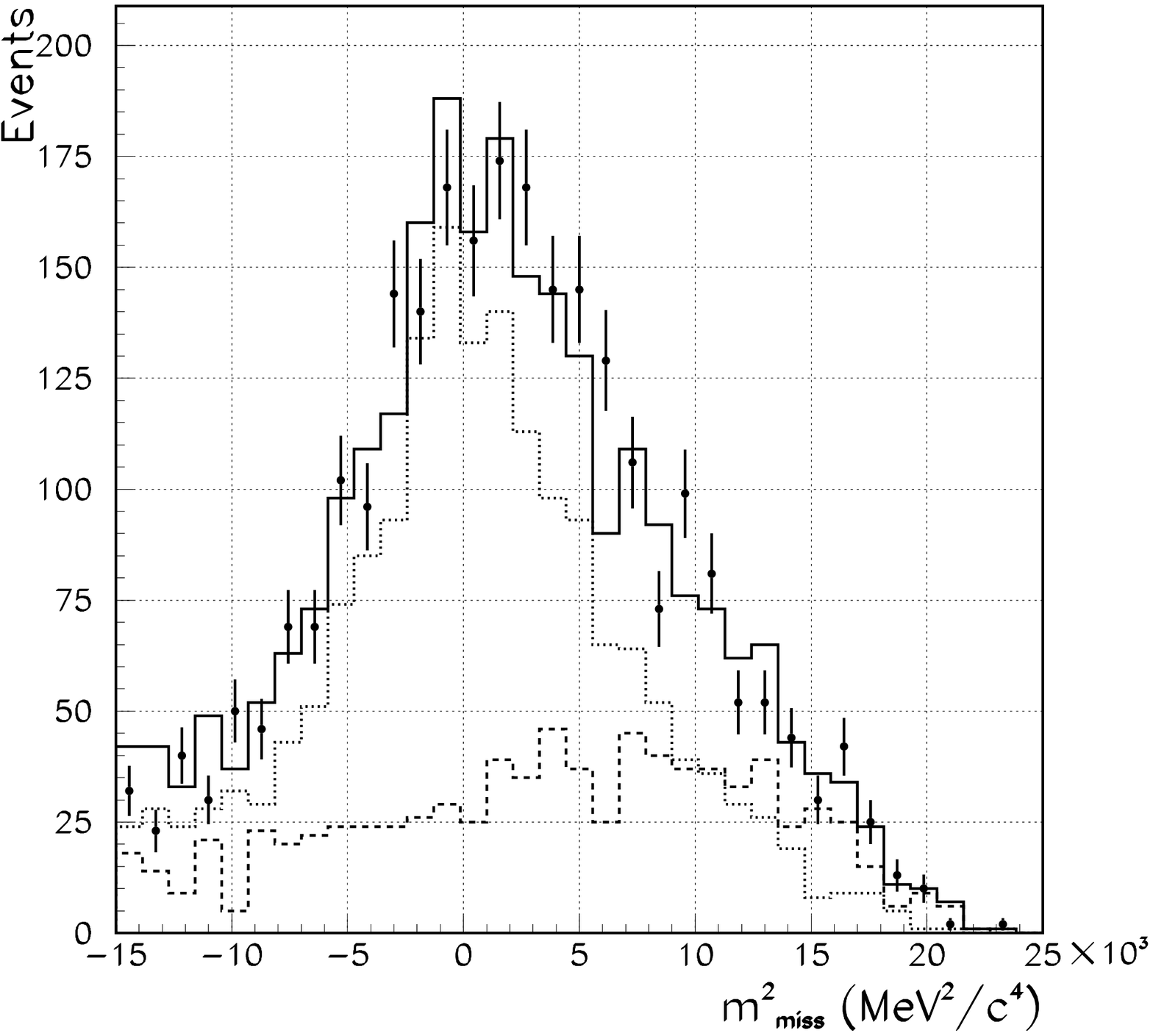} 
\includegraphics[width=7.0cm]{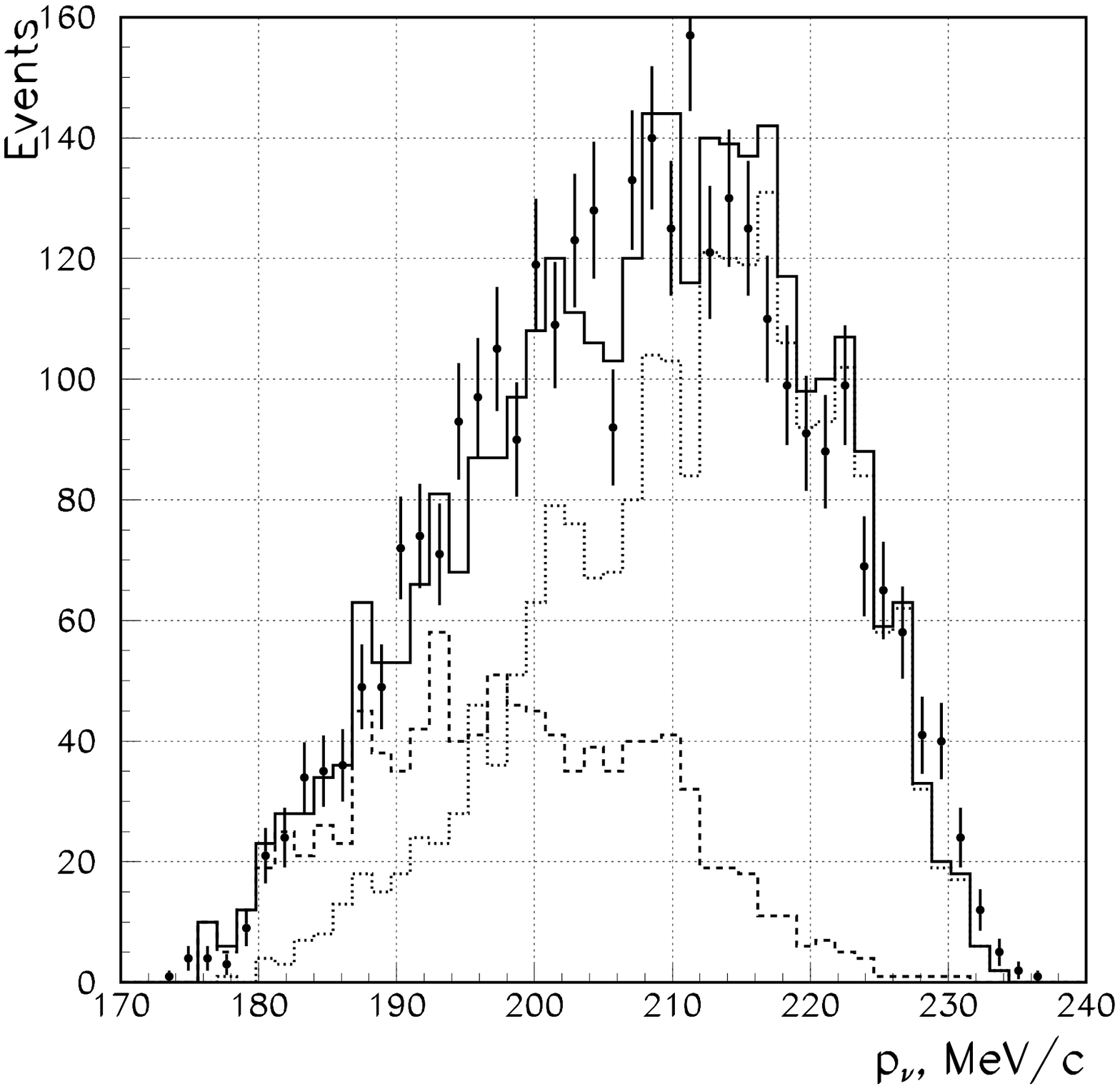} 
\caption{The spectra of the  neutrino missing mass (left) and 
momentum (right) for 
$\kmng$ selection. The black dots with error bars show the experimental 
data. MC simulation: dotted  line~--~$K_{\mu2\gamma}$, 
dashed line~--~$K_{\mu3}$, solid line~--~$K_{\mu2\gamma} + K_{\mu3}$.}
\label{fig:kmng}
\end{figure}

The second stage for $\kmng$ selected one-photon events with $E_{\gamma}>50$ MeV 
and comprised 
three major cuts: a constraint on the  neutrino missing mass 
$-0.7 \times 10^4 < M^2_{miss} < 1.5 \times 10^4$ MeV$^2$/c$^4$, a cut on 
the neutrino 
momentum $p_{\nu} > 200$ MeV/c and a cut on the opening angle between muon and photon 
$\Theta_{\mu\gamma}<90^\circ$. These cuts suppressed  
the $\kmnp$  by a factor of 70, while sustaining a 
$\kmng$ loss by a factor of 2.
Fig.~\ref{fig:kmng} 
shows the spectra of $M^2_{miss}$ and $p_\nu$ for 
both Monte Carlo~(MC)
($\kmnp$ and $\kmng$) and experimental data.

The background fractions for extracted $\kmnp$ and $\kmng$ events were 
estimated 
using 
experimental spectra along with the MC simulation data. For $\kmnp$ the major background 
contributions come from CsI(Tl) accidental hits, $K_{e3}$ and $K_{\pi 2}$. All these  
backgrounds do not induce spurious asymmetry and  only dilute the sensitivity to 
$P_T$.
The total background fraction for $\kmnp$ was estimated to be $\leq 16.0$\%. 

In the case of $\kmng$ the situation is much worse due to the predominant 
background of $\kmnp$ 
events with one photon escaping CsI(Tl) detector. Such events almost completely 
mimic 
$\kmng$ kinematics, so they cannot be suppressed without a considerable loss of 
useful 
$\kmng$ events. The optimized $\kmng$ cuts reduced the 
background fraction 
from $\kmnp$ to about 17\%. 
The background $\kmnp$ events might have non-zero $P_T$ thereby 
inducing spurious transverse asymmetry. Fortunately,  
we measure $P_T$ in $\kmnp$  with higher sensitivity  in the same 
experiment and can reliably estimate this effect.  Therefore, we can safely assume 
no spurious 
contribution to $P_T$ from $\kmnp$ background. The second major source of 
background is  accidental photons  in the   CsI(Tl) detector. It was suppressed 
to the level of $\leq 8$\%
by requiring a photon  energy  threshold of 50 MeV and a coincidence between a
signal from charged particle and a photon signal in the CsI within a window 
$\pm 15$ ns.  
Other background 
modes are suppressed by the $\kmng$-specific cuts to a negligible level. The 
total background 
fraction for $\kmng$ sample was estimated to be $\leq 25$\%.

\begin{figure}[htb]
\includegraphics[width=7.0cm]{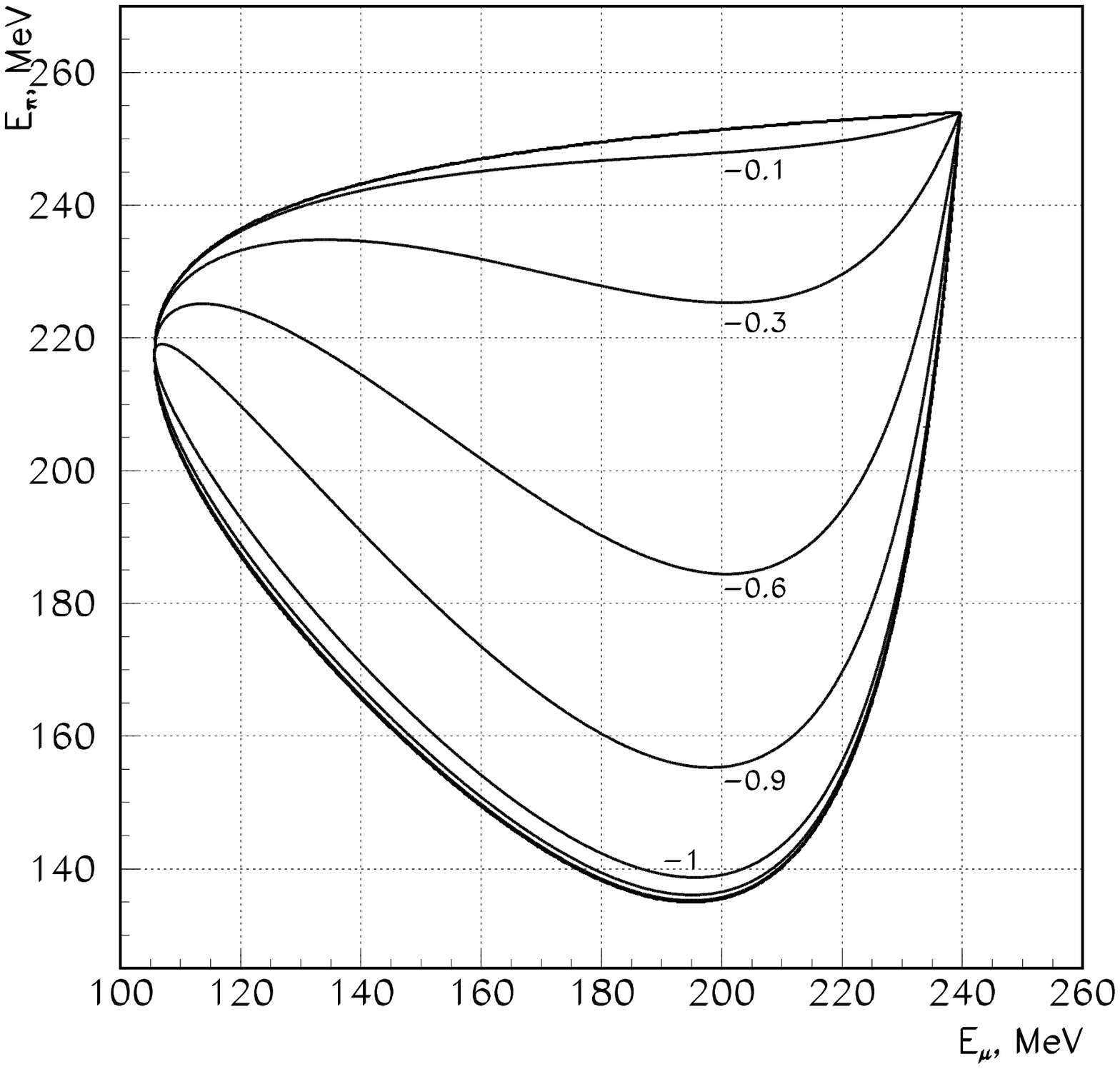} 
\includegraphics[width=7.0cm]{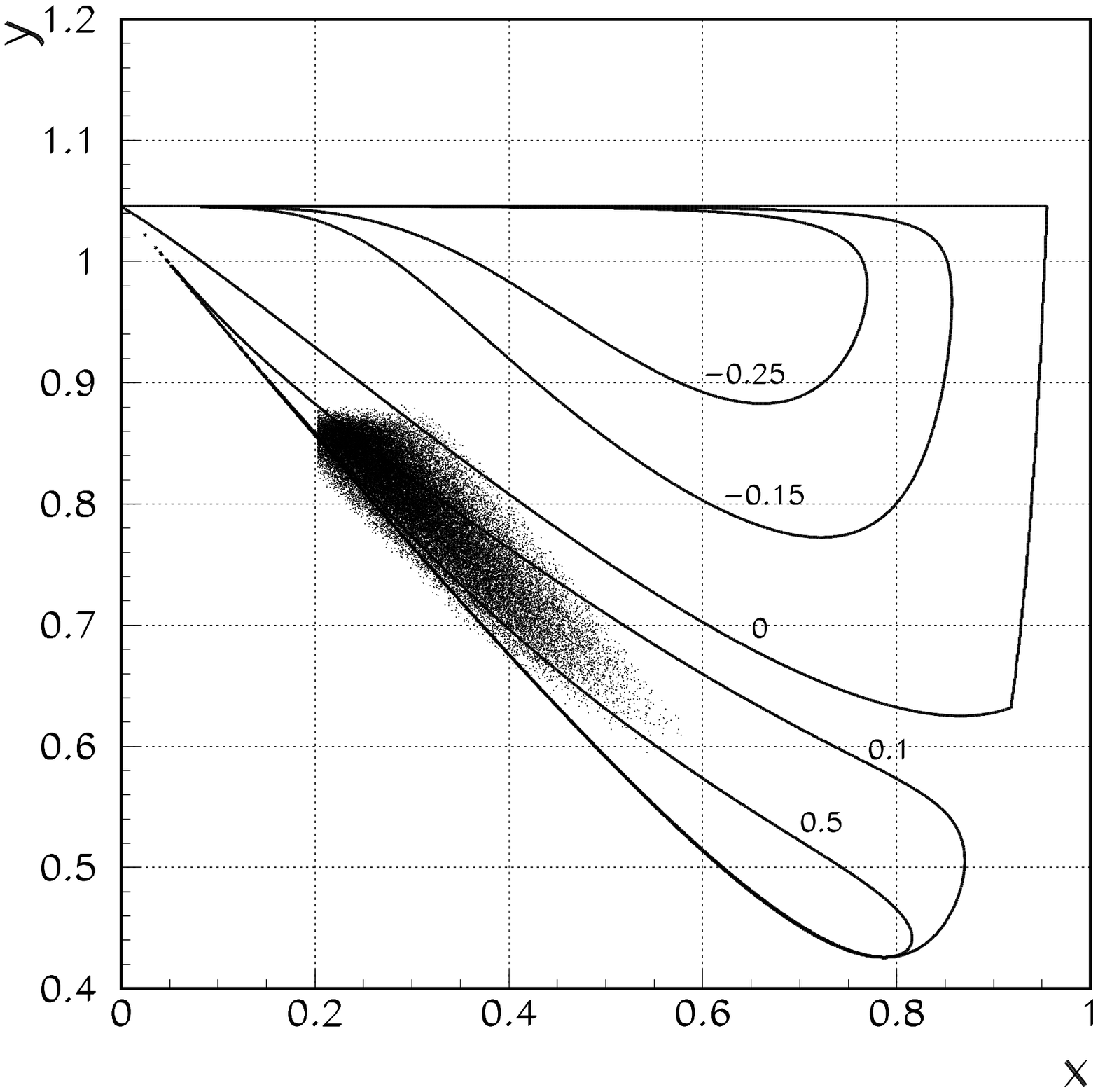} 
\caption{Contour lines of the normal polarization over the 
Dalitz plot: $\kmnp$ (left) and 
$\kmng$ (right). The dots represent the $\kmng$ experimental data.}
\label{fig:pn}
\end{figure}

The valuable part of the asymmetry analysis includes the extraction of the value 
of the normal asymmetry $A_N$ which is proportional to the T-even muon 
polarization, i.e. the in-plane component of the muon polarization  normal to the
muon momentum. It
can be measured 
by selecting events with $\pi^0$ (or photon for $\kmng$) moving into the left or 
right 
hemisphere with respect to the median plane of the given magnet sector.
The theoretical calculations indicate that the normal polarizations 
for $\kmnp$~\cite{pn_kmu3} 
and $\kmng$~\cite{chen} have opposite signs for the kinematic region 
where selected events are located, as  shown in Fig.~\ref{fig:pn}.
For one-photon $\kmnp$ events, we obtained 
$A_N(\kmnp(1\gamma))=(-3.87 \pm 0.06)\times 10^{-2}$
while for $\kmng$ we have $A_N(\kmng)=(+3.67 \pm 0.44)\times 10^{-2}$. 
This gives  us  
a robust evidence of the sufficiently pure $\kmnp$ and $\kmng$ selection. 
The study of the dependence 
of the normal asymmetry on the energy of $\pi^0$ shows a sound agreement with 
the theoretical 
calculation. By comparing the values of the normal asymmetry and normal 
polarization 
obtained using Monte Carlo simulation using the relation $A_N=\alpha\times P_N$, we 
are able to extract the value of the polarimeter
analyzing power  $\alpha=0.289 \pm 0.015$. 

The main systematic uncertainties in  $P_T$ come  from magnetic field 
rotation, detector 
components misalignment, CsI(Tl) accidental background, beam profile asymmetry 
and decay plane 
rotation. The important point is that the sources of systematics are mostly 
the same 
for both measured decays, so we can estimate the systematic contributions using 
a large sample 
of $\kmnp$ data and then use the same error value for $\kmng$. Most of the 
systematic effects are 
canceled by the unique features of the  set-up (azimuthal symmetry, 
double ratio, etc).
We evaluated the contributions from all relevant sources using the MC simulation 
data and 
comparing them with the experimental distributions~\cite{kek1}. Overall systematic 
error 
was estimated to be 
$\delta P_T=0.92\times 10^{-3}$ which is well below the level of the 
statistical error.

\section{Results}

After the analysis of the entire data set collected in the period 1996-2000 we obtained the 
results for both $\kmnp$ and $\kmng$. For $\kmnp$ we selected
$6.3\times 10^6$ 
one- and two-photon forward/backward events and using the angular attenuation factor value 
extracted from MC simulation ($f=0.72 \div 0.77$ for $2\gamma$ events and $f=0.56 \div 0.66$ for 
$1\gamma$ events) the preliminary result  
$P_T=(-1.12 \pm 2.17(stat) \pm 0.92(syst)) \times 10^{-3}$ is obtained. 
Using the relation $P_T=\mbox{Im}(\xi)\cdot\Phi$ (where $\Phi$ is a kinematic factor, evaluated 
from MC simulation $\Phi \sim 0.2 \div 0.3$) we get 
$\mbox{Im}(\xi)=(-0.28 \pm 0.69(stat) \pm 0.30(syst))\times 10^{-2}$ \cite{kek2}. 
The dependences of the transverse asymmetry on the beam cycle, $\pi^0$ energy and magnet 
sector 
number indicate no systematic irregularity and thus confirm the robustness of our systematics 
study. The results indicate no evidence of T-violation in $\kmnp$ and can be interpreted as 
limits on the measured quantities: $|P_T|<4.3\times 10^{-3}$ at 90\% c.l. and 
$|\mbox{Im}(\xi)|<1.3 \times 10^{-2}$ at 90\% c.l. 

We have performed  the first measurement of $P_T$ in the 
$\kmng$ decay (it is also the first $\kmng$ 
measurement  below $K_{\pi 2}$ peak). The result obtained for 1996-1998 data
was published in~\cite{anisimovsky}. Here we present the result of the analysis 
of the whole data set of $1.88\times 10^5$ forward/backward events. 
We obtained  the value 
$P_T=(-0.14 \pm 1.44(stat) \pm 0.10(syst))\times 10^{-2}$  with the 
 evaluated  attenuation factor of  $f=0.80 \pm 0.03$.  
Similarly to the case of $\kmnp$, we see no indication of T-violation in 
this  decay and we can put the 
limit 
$|P_T|<2.4 \times 10^{-2}$ (90\% c.l.) 

\section{Conclusion}

We have performed the new measurement of T-violating muon polarization in two 
decays $\kmnp$ and 
$\kmng$ for which several non-standard models predict non-zero $P_T$ values. 
For $\kmng$ decay 
our result is the first one. At the current  level of the experimental
 sensitivity, we see 
no evidence for T-violation 
and our results allowed us to impose constraints on several 
non-standard models: three Higgs doublet 
model (the most stringent experimental constraint), SUSY with squark mixing, 
SUSY with R-parity 
violation, leptoquark models, left-right symmetric models (see, for example,  
 Ref.~\cite{Bezrukov:2003jc}).
Much higher  statistical sensitivity to $P_T$  of $\leq 10^{-4}$  can be reached 
in a proposed  experiment~\cite{Kudenko:yk} at the high intensity low 
energy separated  kaon beam 
at J-PARC~\cite{jhf}. In addition, the  $P_N$  values in $K_{\mu3}$ and 
 $K_{\mu2\gamma}$ can be measured with high accuracy  in this experiment that
 provides a new sensitive method for determination of the  kaon form factor 
 values  in these decays~\cite{bezrukov}.

\end{document}